\documentstyle[epsf,12pt]{article}
\begin{document}


\begin{center}
{\large {\bf The range of the contact interactions and the kinetics
of the Go models of proteins }}
\end{center}

\begin{center}
{\bf
{\sc Marek Cieplak$^1$, and Trinh Xuan Hoang$^2$
}}
\end{center}

\vspace*{1.5cm}
\begin{center}
$^1$Institute of Physics, Polish Academy of Sciences, 02-668
Warsaw, Poland\\
$^2$The Abdus Salam International Center for Theoretical Physics,\\
Strada Costiera 11, 34100 Trieste, Italy\\ 
\end{center}

\vskip 40pt


\vskip 40 pt
\begin{abstract}
We consider two types of Go models of a protein (crambin) and study
their kinetics through molecular dynamics simulations.
In the first model, the residue -- residue contact interactions
are selected based on a cutoff distance, $R_c$, between the
C$_\alpha$ atoms. The folding times
depend on the value of $R_c$ strongly and non-monotonically due
to the interplay between frustration and the free energy barrier for
folding.
This indicates a need for a
physically determined set of native contacts that takes
into account all the residual atoms.  
This can be accomplished by considering the van der Waals radii
of the atoms and checking if 
they are found within a proper range of the van der Waals
attraction.
In the second model,
non-native attractive contacts are added to the
system. This leads to bad foldability. However, for a small
number of such extra contacts there is a slight acceleration
in the kinetics of folding.
\end{abstract}

\vskip 30 pt
\noindent {\bf keywords: protein folding; folding rate;
molecular dynamics; chirality}

\newpage
\begin{center}
{\bf INTRODUCTION}
\end{center}

All-atom molecular dynamics (MD) simulations are not yet
adequate tools to study kinetics of protein folding.
Due to a large number of degrees of freedom involved
the accessible time scales in MD are orders of magnitude too short
compared to the experimental folding times which last
for a millisecond or longer \cite{Ruczinski}. Thus theoretical
studies of the kinetics must involve simplified models such
as those in which each amino acid is represented by a single bead that is
located at a position of the C$_\alpha$ atom. This kind of a
drastic simplification in turn requires a design of effective
interactions between the beads.
A determination of such interactions itself  turns out not to be
a simple task if one requires preservation of the chemical identities
of the amino acids. Go models are
a class of simplified systems that do not take into
account the sequential specificity of the protein
and yet have proved to be fairly
realistic in terms of the kinetic properties of proteins
\cite{Goabe,Stakada}.
These models are built based merely on the knowledge
of the experimentally determined native state. The interactions
between the beads are made attractive
if the pair of residues form a native contact and
repulsive otherwise (to take the excluded volume into account). \\


The selection of what pair of residues is considered as forming a native
contact is a basic ingredient of a Go-type model.
A common procedure is to consider the distances, $r$'s, between
the C$^{\alpha}$ atoms in the native state, and to adopt a certain
cut-off distance $R_c$ such that an $r$ smaller than $R_c$ gives rise
to a native contact. The value of $R_c$ has been chosen in the
literature quite arbitrarily, between 6.5\AA and 8.5\AA.
Though in principle there should be some optimal choice of $R_c$ that reflects
the size and structural details of the amino acids, the simple expectation
is that the folding kinetics is not too sensitive to the
specific value of $R_c$ within a reasonable range.
In this paper, we examine in details the dependence of folding kinetics
on $R_c$ and show that this dependence is
actually rather strong and quite non-trivial. As an illustration
we consider the Go-type models of crambin
and find that the fastest folding time, as determined under
optimal folding conditions, depends on $R_c$ non-monotonically,
although the non-monotonicity is restricted to a rather narrow range.
Generally, however, the bigger the
$R_c$, the shorter the folding time and the bigger the stability
because more and more interactions favor moving into the native
state.\\

A more physical way \cite{Prion} to determine the native contacts
involves taking all pairs of the non-hydrogen
atoms in the two amino acids, assigning the van der Waals radii \cite{Tsai}
to them and checking if there is an overlap. The criterion
for the overlap takes into account some softness in the potential
and assigns a factor of 1.244 (the inflection point in the
Lennard Jones potential) to the sum of atomic radii. This
factor is again somewhat subjective but the whole procedure
involves considering the actual sizes of the amino acids
and the corresponding Go model will serve as a "realistic"
reference system here. For crambin, the distribution of the
resulting 137 contact distances is shown in Figure 1 and the structure
file was taken from the PDB \cite{PDB}. The contact distances
range from 4.1 to 9.5 $\AA$. The $R_c$-based criterion would involve
the van der Waals-determined contacts up to the distance of
$R_c$ but also many additional pairs which are upgraded to
contacts artificially. On comparing kinetics of the reference system
to those corresponding to various $R_c$s one can find a value
for which there is a close resemblance. For crambin, we find that
the corresponding equivalent $R_c$ is near $7.5 \AA$.\\

The next issue which we study here is how the folding kinetics
are affected when attractive interactions are added to the non-native
contacts.
Recently, Plotkin \cite{Plotkin} has argued
that, as he put it in the title of his paper,
"a little frustration sometimes helps", i.e. 
adding some small noise to non-native interactions
may actually accelerate the kinetics of folding
up to several times.
In Plotkin's model the non-native contacts are assigned with a random
energy with some small variance.
The models we studied are in a similar spirit but in our models
non-native attractions are assigned to only few non-native contacts while
their energies are the same as those of the native ones.
This is done by first
generating the reference (van der Waals-based) Go model and then
by randomly selecting $n_a$ extra pairs with the
sequence distance of at least 4. These extra pairs are endowed
with the repulsive potential in the reference model and now
they acquire a potential which is attractive.
In agreement with what proposed by Plotkin,
the acceleration of the folding kinetics is also observed
in our models but we find it to be rather weak and it happens
only when $n_a$ is small.
After some threshold, the basic effect of increasing the $n_a$
is to  turn the systems into poorer and poorer folders.\\

\begin{center}
{\bf MODELS AND METHODS}
\end{center}

\underline{The Hamiltonian}\\

An input for the construction of the Go model is a PDB file \cite{PDB}
with the coordinates
of all atoms in the native conformation. The coordinates are used
to determine the length related parameters of the model.\\

There are many variants of the Go models depending, for instance,
on the choice of the functional form of the attractive potentials
in the native contacts. The 10-12 potentials
(with the $r^{-12}$ repulsion and $r^{-10}$ attraction)
are a common choice,
see for instance \cite{Nymeyer,Prion}. Another, used here, is
based on the Lennard Jones form. We follow the procedure as described
in our previous papers \cite{Hoang,Hoang1,optimal,stretch1,stretch2}
and especially in \cite{universal}.
The Hamiltonian
consists of the kinetic energy and of the potential energy,
$E_p(\{\bf r_i\})$,
which is given by
\begin{equation}
E_p(\{{\bf r_i}\})\;=\; V^{BB} \;+\;V^{NAT} \;+\;V^{NON} \;+\;V^{CHIR} \;\;.
\end{equation}
The first term, $V^{BB}$ is the harmonic potential
\begin{equation}
V^{BB} = \sum_{i=1}^{N-1} \frac{1}{2}k (r_{i,i+1} - d_0)^2 \;\;,
\end{equation}
which tethers consecutive beads at the equilibrium bond length,
$d_0$, of 3.8{\AA}. Here,
$r_{i,i+1}=|{\bf r}_i - {\bf r}_{i+1}|$ is the distance between
the consecutive beads and $k=100 \epsilon /$\AA$^2$,
where $\epsilon$ is the energy scale characterizing the native contacts.\\

$V^{NAT}$ corresponds to the Lennard-Jones interactions
in the native contacts and is given by
\begin{equation}
V^{NAT}_{6-12} =
\sum_{i<j}^{NAT}4\epsilon \left[ \left( \frac{\sigma_{ij}}{r_{ij}}
\right)^{12}-\left(\frac{\sigma_{ij}}{r_{ij}}\right)^6\right],
\end{equation}
where the sum is taken over all native contacts and $\epsilon$
is common to all contacts.
The parameters
$\sigma_{ij}$ are chosen so that each contact in the native
structure is stabilized at the minimum of the potential,
and $\sigma \equiv 5$\AA\ is a typical value.
For each pair of interacting amino acids, the two potentials
have a minimum energy of $-\epsilon$. 
The non-native interactions, $V^{NON}$, are purely repulsive and are
necessary to reduce the effects of entanglements.
They are taken as
the repulsive part of the Lennard-Jones potential that corresponds
to the  minimum occurring at 5{\AA}. This potential is truncated
at the minimum and shifted upward so that it reaches zero energy
at the point of truncation.\\

The last term in the Hamiltonian, $V^{CHIR}$,
favors the native sense of the chirality
at each location along the backbone.
A chirality of residue $i$ is defined as
\begin{equation}
C_i = \frac{\left( {\bf v}_{i-1} \times {\bf v}_{i} \right)
\cdot {\bf v}_{i+1}}{d_0^3},
\end{equation}
where ${\bf v}_{i}={\bf r}_{i+1} - {\bf r}_{i}$.
A positive $C_i$ corresponds to right-handed chirality. Otherwise the
chirality is left-handed.
$V^{CHIR}$ is given phenomenologically \cite{universal} by
\begin{equation}
V^{CHIR} =
\sum_{i=2}^{N-2}\frac{1}{2} \; \kappa \; C_i^2 \; \Theta ( - C_i^{NAT} ),
\end{equation}
where $\Theta$ is the step function (1 for positive arguments
and zero otherwise),
$C_i^{NAT}$ is the chirality
of residue $i$ in the native conformation,
and $\kappa$ is taken to be equal to $\epsilon$.
The role of $V^{CHIR}$ is primarily to punish energetically any
deviations from the non-native sense of chirality.\\

The attractive non-native contacts, when built in,
are given by the Lennard Jones attraction with
the minimum at 5 $\AA$.\\

\underline{The time evolution}\\

The time evolution of unfolded conformations to the native state
is simulated by MD as described in \cite{Hoang,Hoang1}.
The beads representing the amino acids
are coupled to Langevin noise and damping terms that provide
thermostating at a temperature $T$.
The equations of motion for each bead are
\begin{equation}
m\ddot{{\bf r}} = -\gamma \dot{{\bf r}} + F_c + \Gamma \ \ ,
\end{equation}
where $m$ is the mass of the amino acids represented by each bead.
The specificity of masses has turned out to be irrelevant
for kinetics \cite{stretch1} and it is sufficient to to consider
masses that are uniform and equal to the average amino acidic mass.
$F_c$ is the net force due to the molecular potentials
and external forces,
$\gamma$ is the damping constant, and
$\Gamma$ is a Gaussian noise term with dispersion
$\sqrt{2\gamma k_B T}$.
For both kinds of the contact potentials, time is measured in units of
$\tau \equiv \sqrt{m \sigma^2 / \epsilon}$, where $\sigma$ is 5{\AA}.
This corresponds to the
characteristic period of undamped oscillations at the bottom
of a typical Lennard-Jones potential.
For the average amino acidic mass and $\epsilon$ of order 4kcal/mol,
$\tau$ is of order 3$ps$.
We have found that the folding times, $t_{fold}$ depend on
$\gamma$ linearly so going to more realistic larger values of
viscosity, as controlled by $\gamma$ involves simple
rescaling.
The equations of motion are
solved by means of the fifth order Gear predictor-corrector
algorithm \cite{Gear} with a time step of $0.005\tau$.
\\

The folding time is calculated as the median first passage time,
and is estimated based on at least 201 trajectories.
$T_{min}$ is defined as a temperature
at which $t_{fold}$ has a minimum value when plotted vs. $T$.
For short proteins, the U-shaped dependence of $t_{fold}$ on $T$
may be very broad and then $T_{min}$ is defined as the
position of the center of the U-shaped curve.
The simplified criterion for an arrival in the native conformation
to be declared
is based on a simplified approach in which
a protein is considered folded if all beads that form a native
contact are within the cutoff distance of $1.5 \sigma _{ij}$.\\

The stability temperature $T_f$ is determined through the
nearly equilibrium calculation of the probability that the protein
has all of its native contacts established. $T_f$ is the temperature at
which this probability crosses $\frac{1}{2}$.  The calculation is based
on at least 5 long trajectories that start in the native state.
A protein is considered as a good folder if $T_f$ is found within a range
of temperature at which folding is relatively fast, i.e. preferably
close to $T_{min}$ \cite{prl}.
\\

\begin{center}
{\bf DEPENDENCE OF FOLDING ON R$_c$ }
\end{center}

In our studies of the effects of the cut-off distance $R_c$
in the Go models of crambin (the PDB code is 1crn),
we consider the range from 6 $\AA$ to 11 $\AA$ in which
the number of native contacts varies between 75 and 379
(if one excludes the peptide bond interactions between
the successive beads then there is a total of 990 possible
contacts since the sequence length is 46). The 6$\AA$ case
is borderline because it corresponds to one amino acid
(the 36'th along the sequence) which does not belong to
any contact and the whole structure is stabilised by the
remaining contacts.

Figure 2 shows the median values of $t_{fold}$ as a function
of temperature, $T$. The top panel is for the reference system,
considered to be the "true" model of crambin
whereas the bottom panel is for three values of $R_c$. In each case,
the $T$-dependence has the common U-shaped form but the shape
itself depends on $R_c$ strongly. It becomes very broad for
$R_c$ at and above 8.75 $\AA$. This suggests that the nature of the
model changes profoundly around 8.75 $\AA$. Furthermore,
as indicated by the top panel of Figure 2, $R_c$ of near 7.5 $\AA$
provides the closest, but by no means perfect, representation
of the true model. 7.5 $\AA$ can be then considered as an
equivalent value but this quantity is protein dependent (7.5 $\AA$
appears too big for the domain of titin \cite{universal}).\\

It is expected that adding more and more contacts that tie
the native structure stronger and stronger should have a dual effect:
one is that the thermodynamic stability should get enhanced and the
other is that the kinetic links with the native state multiply
which should accelerate the folding.
Figure 3 suggests, however, that this picture, though generally
correct, is more subtle: the fastest folding time, as determined at
$T_{min}$, is non-monotonic around $R_c$ of 8.75 $\AA$ even though
the number of contacts, $n_{nat}$ remains monotonic throughout.
This can be explained by the argument that while the free energy
barrier for folding decreases on increasing $R_c$
the effects of frustration also get enhanced. The
range of $R_c$ where $t_{fold}$ increases corresponds to the
situation where the increment in frustration is the winning factor.
\\

Whatever the value of $R_c$ up tp 11 $\AA$, $T_f$  stays
in the temperature range corresponding to fast folding, i.e.
on coolling down, the sequence acquires appreciable probability
of staying near the native basin before the glassy effects
set in. This is seen in Figure 2 and also in Figure 4.
The latter figure shows the values of $T_f$, $T_{min}$, and $T_{g2}$
as a function of $R_c$.
$T_{g2}$ is defined as the temperature at which the folding time is
twice as high as the fastest time, on the low temperature side
of the U-curve. For broad U-curves, $T_{g2}$ is a better measure
of when the glassy effects set in than $T_{min}$.
Each of these quantities grow monotonically with $R_c$ (and become
straighter when plotted vs. $n_{nat}$) and $T_f$ stays between
(or near) $T_{g2}$ and $T_{min}$ which indicates good foldability.\\

Interestingly, $T_f$ of the reference system agrees with
a $T_f$ obtained for $R_c$ between 7 and 7.5 $\AA$ whereas
$T_{min}$ of the reference system is equivalent to
$R_c$ slightly higher than 6 $\AA$ which suggests that
the true system cannot really be represented by any uniform
cutoff value in the contact range.\\

Figure 5 shows the specific heat, as obtained by the weighted histogram
method \cite{Ferrenberg}, which provides an additional
characterization of the equilibrium properties. The positions of
the maxima grow with $R_c$ and so do the maximal values.
The temperature width, however, remains more or less constant.
Note that even though $R_c$ of 7.5 $\AA$ provides a reasonable
fit to the plots of the folding time vs. $T$ for the reference
system, the specific heat is off more noticeably
but the peak position is adequate.
It should be noted that the peak positions in the specific heat are higher
than both $T_f$ and $T_{min}$ for all cases.
The emergence of a single maximum in the specific heat is a signature
of an equilibrium folding transition
when the affinity towards the
native basin starts to dominate the free energy.
The fact that $T_f$ is smaller than the temperature of the specific heat's
peak indicates that below the folding transition temperature
the entropy associated with the chain is still significant
and not all of the native contacts are established. Complete folding
seems to be more accurately associated with $T_f$, which is defined in
reference to a single micro state -- the native conformation.
\\

\begin{center}
{\bf THE EFFECTS OF NON-NATIVE ATTRACTIVE CONTACTS}
\end{center}

We now consider the second model in which $n_a$ attractive
non-native contacts are added to the reference Go model
of crambin. We consider just one realization for the contact
addition for each $n_a$ and a system corresponding to a larger
value of $n_a$ incorporates contacts generated at any
smaller value of $n_a$.\\

We have checked that for $n_a \leq 90$ the native
state does not change its nature and it
remains accessible kinetically.
Its stability, however, decreases with an increasing $n_a$.
Figure 6 shows the U-shaped curves of the folding
time vs. $T$. In contrast to what happens in the previous
model, the curves become narrower and narrower on adding
contacts. Furthermore, $T_f$ drifts towards lower and lower
temperatures, eventually leaving the region of fast folding.
For $n_a$ greater than $\sim$ 60, the systems are bad folders.\\

The fastest folding time is shown in Figure 7 as a function of $n_a$.
It is interesting to note that for small values of $n_a$
there is a weak decrease
in $t_{fold}$ compared with the case of $n_a$ of 0.
Though this decrease is small, it is in a qualitative
agreement with Plotkin's argument which suggests that
non-native interactions involved in the collapsed phase
may lower the free energy barrier for folding \cite{Plotkin}.
Beyond $n_a$ of about 25, there is a steady and nonlinear growth
of $t_{fold}$ with $n_a$.\\

Figure 8 shows $T_f$, $T_{min}$, and $T_{g2}$ and a function of
$n_a$. Their relative positioning indicates a flow from
good to bad folding properties and bad native stability.
$T_f$ becomes smaller than $T_{g2}$ at $n_a$ of about 50.
For a sufficiently large $n_a$, $T_f$ is expected to disappear.\\

The specific heat for several values of $n_a$ is shown in Figure 9.
In contrast to the picture shown in Figure 4,
which also deals with a growing number of contacts,
an increase in $n_a$ results in a decrease
in the specific heat's maximum and in shifting it towards lower
temperatures. A remarkable change is observed at $n_a$ of 75
and higher, where the specific heat curve acquires a double humped shape.
These two peaks can be interpreted as corresponding
to two transitions: of folding and of collapse with the former
appearing at the lower temperature.
The emergence of these two transitions
indicates conditions of bad foldability.
Notice also that in accordance with what was discussed in the previous
section, $T_f$ is always smaller than the temperature
corresponding to the peak in the specific heat (the smaller peak
if there are two).
\\

\begin{center}
{\bf CONCLUSIONS}
\end{center}

In this paper, we demonstrated that the selection of the
proper contact range  has a strong effect on the folding kinetics in
Go models and is not at all innoccuous. It is thus important
to have a physical scheme,
such as based on the van der Waals radii of the atoms, that allows
for an amino acid by amino acid determination of whether
they make a native contact or not. Adding non-native
attractive contacts leads eventually to bad foldability but adding
just a few attractive non-native contacts slightly accelerates
the folding process.\\

\newpage
\begin{center}
{\bf ACKNOWLEDGMENTS}
\end{center}
MC thanks the Department of Physics and Astronomy at Rutgers University for
providing him with the computing resources. This research was supported
by Komitet Badan Naukowych Grant 2P03B.  


\newpage

\begin{center}
{\bf FIGURE CAPTIONS}
\end{center}

Figure 1. The distribution of the effective contact lengths in crambin
as determined by the procedure which is based on the van der Waals radii
of the atoms. The shaded region corresponds to the contacts that would
not be included if the cutoff of 7.5 $\AA$ was adopted.\\

Figure 2. The dependence of the folding time on temperature for
various Go models of crambin. The top panel is for the contacts
determined through the criterion that involves the van der Waals
radii of the residual atoms.
The bottom panel is for contacts determined by the cutoff based
criterion. The corresponding values of $R_c$ are indicated.
The crosses on the top panel correspond to $R_c$ of 7.5 $\AA$.
The error bars are of order of the size of the symbol
representing the data  points.
The arrows indicate values of the folding temperature $T_f$.\\

Figure 3. The folding time at $T_{min}$ as a function of
the cutoff distance $R_c$. The error bars are less than
double the circular symbol size.  The stars represent
the numbers of the native contacts corresponding to
a given value of $R_c$.\\

Figure 4. The values of the characteristic temperatures
$T_{min}$ (open squares), $T_f$ (solid circles),
and $T_{g2}$ (triangles) and positions of the
maximum in the specific heat (stars)
for the Go models of crambin for various values of $R_c$.\\

Figure 5. The plots of specific heat as a function of $T$ for
several values of $R_c$, as indicated. The solid line (denoted
by VdW) corresponds to the reference model in which the native
contacts are determined based on the van der Waals radii of
the atoms.
\\

Figure 6. The dependence of the folding time on temperature
for Go models of crambin in which $n_a$ extra non-native
attractive contacts are added. The values of $n_a$ are indicated.
The arrows show values of $T_f$.\\

Figure 7. The folding time as a function of $n_a$.
The dotted line indicates the value corresponding to  $n_a$=0.
\\

Figure 8. Similar to Figure 4 but for the model with the added
non-native attractive contacts.\\

Figure 9. Specific heat as a function of $T$ for the indicated values of
$n_a$.

\onecolumn

\begin{figure}
\vspace*{-0.5cm}
\epsfxsize=7in
\centerline{\epsffile{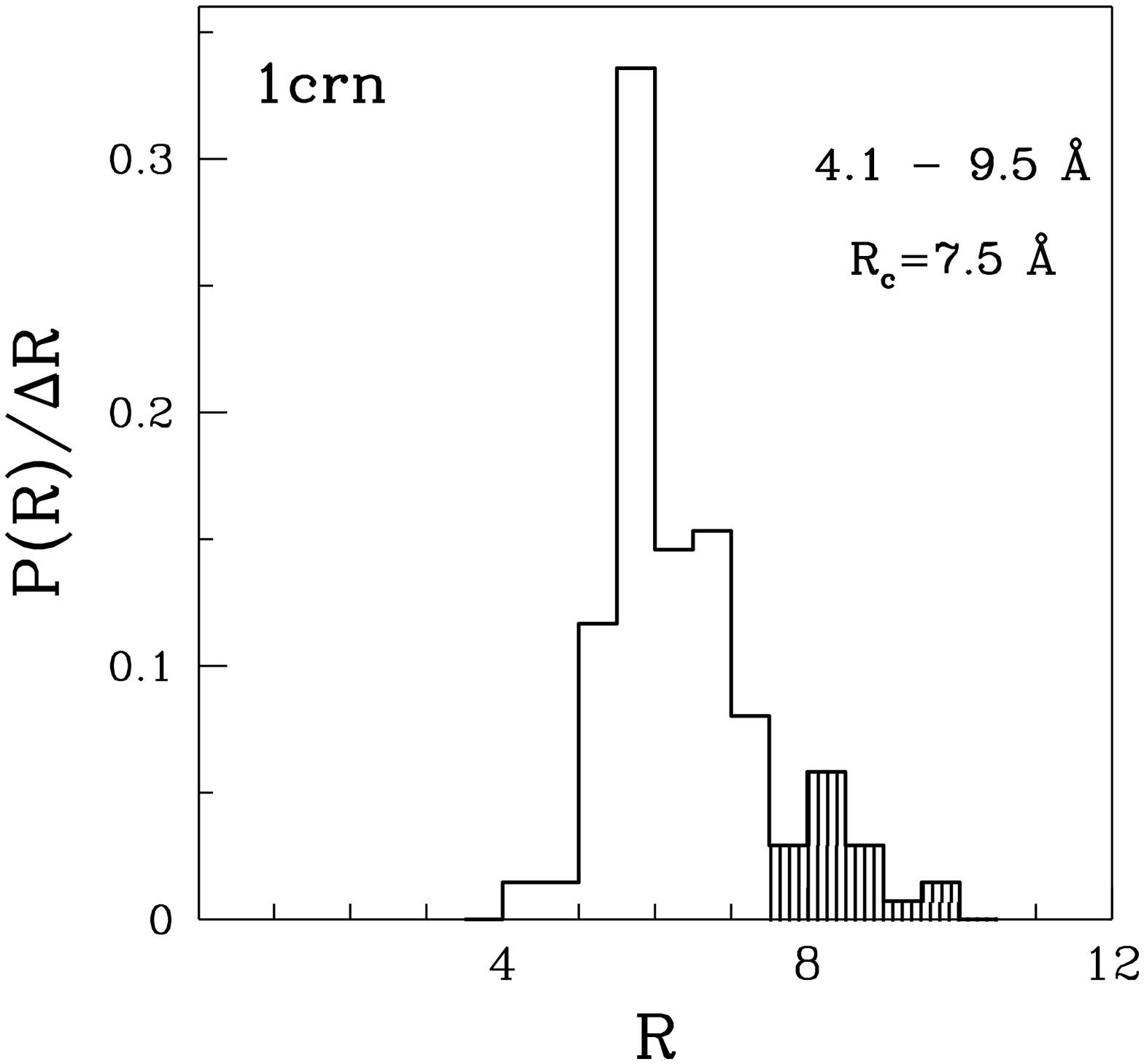}}
\vspace*{0.5cm}
\caption{ }
\end{figure}

\begin{figure}
\vspace*{-0.5cm}
\epsfxsize=7in
\centerline{\epsffile{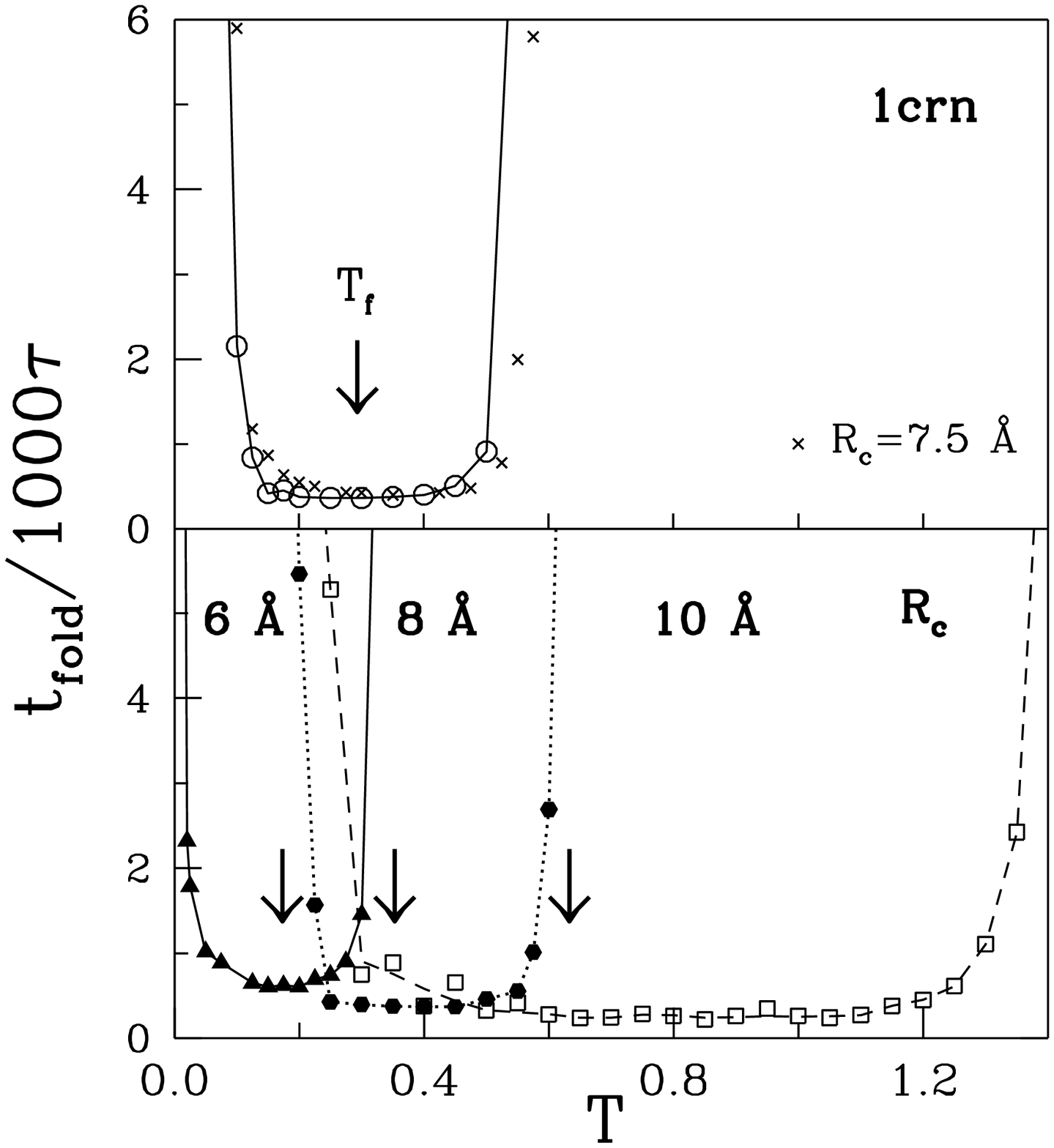}}
\vspace*{0.5cm}
\caption{ }
\end{figure}

\begin{figure}
\vspace*{-0.5cm}
\epsfxsize=7in
\centerline{\epsffile{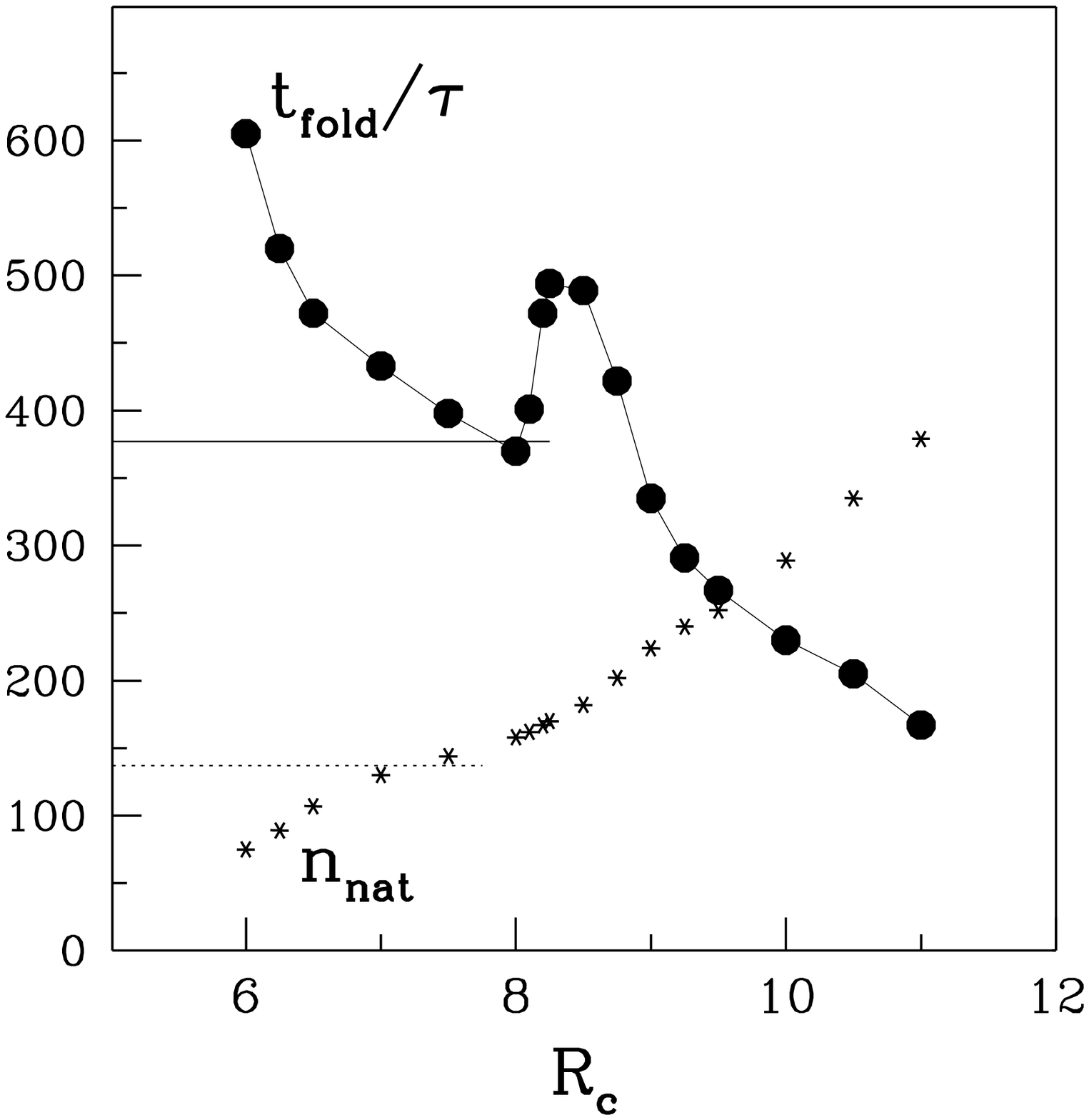}}
\vspace*{0.5cm}
\caption{ }
\end{figure}

\begin{figure}
\vspace*{-0.5cm}
\epsfxsize=7in
\centerline{\epsffile{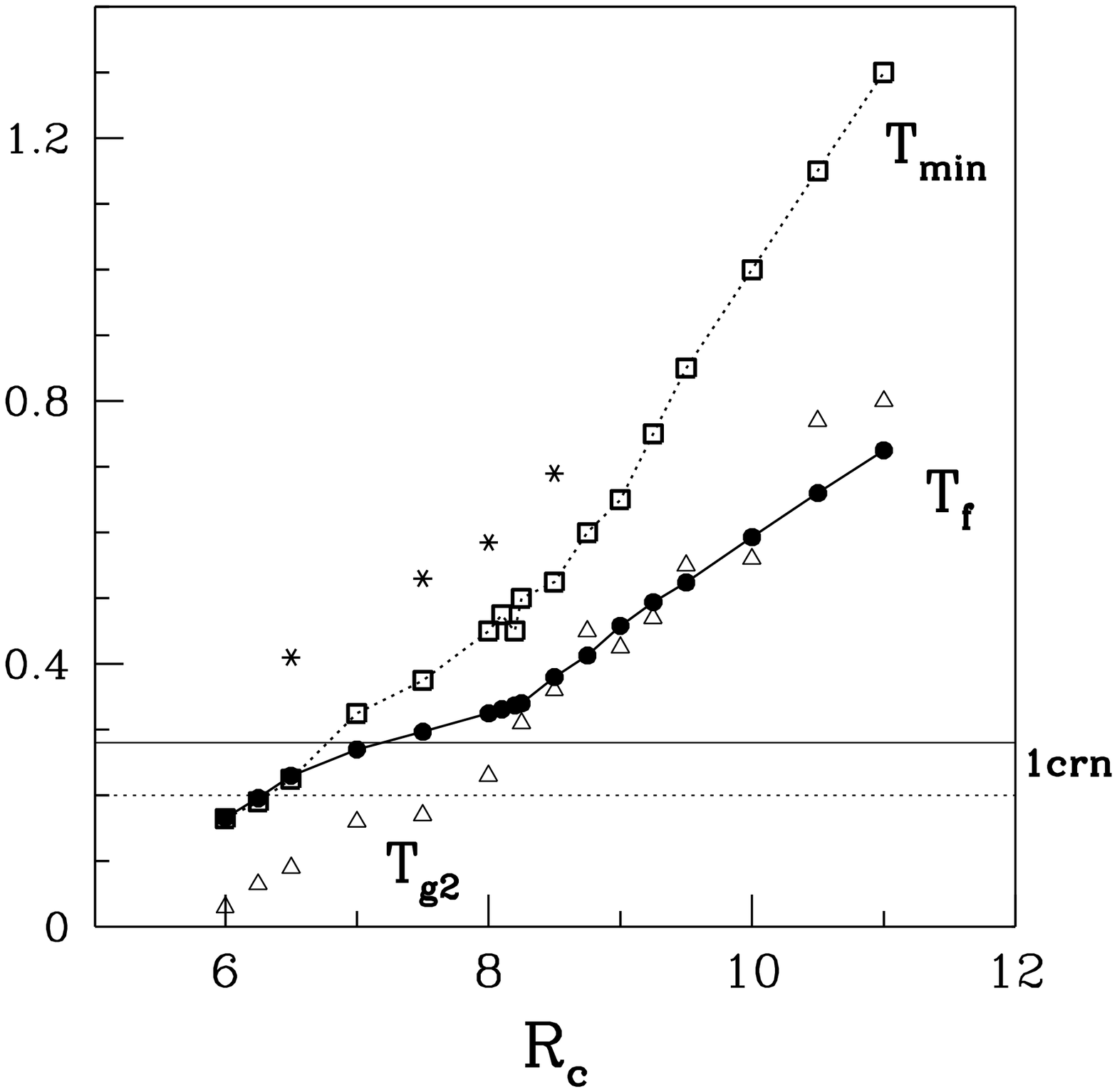}}
\vspace*{0.5cm}
\caption{ }
\end{figure}

\begin{figure}
\vspace*{-0.5cm}
\epsfxsize=7in
\centerline{\epsffile{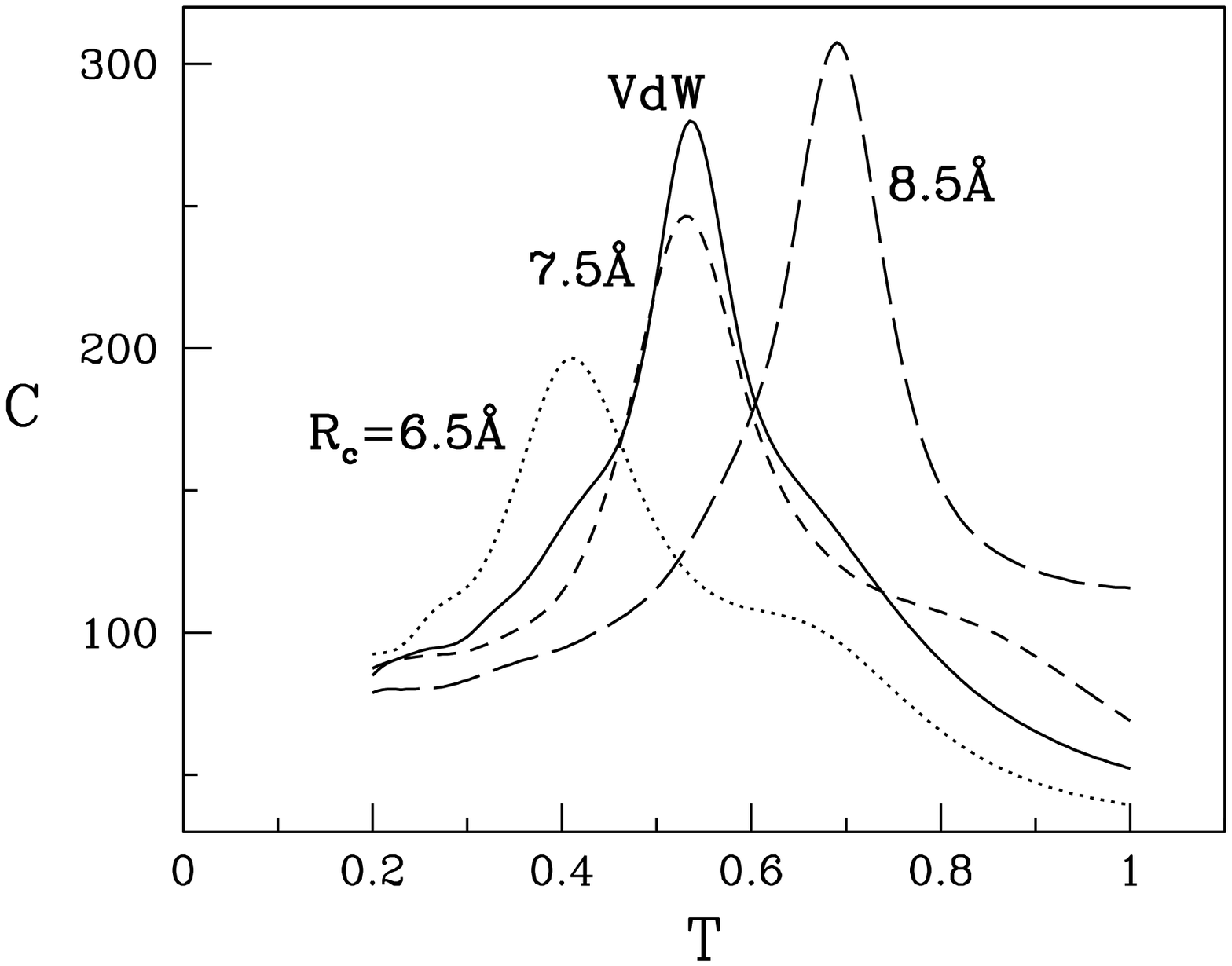}}
\vspace*{0.5cm}
\caption{ }
\end{figure}

\begin{figure}
\vspace*{-0.5cm}
\epsfxsize=7in
\centerline{\epsffile{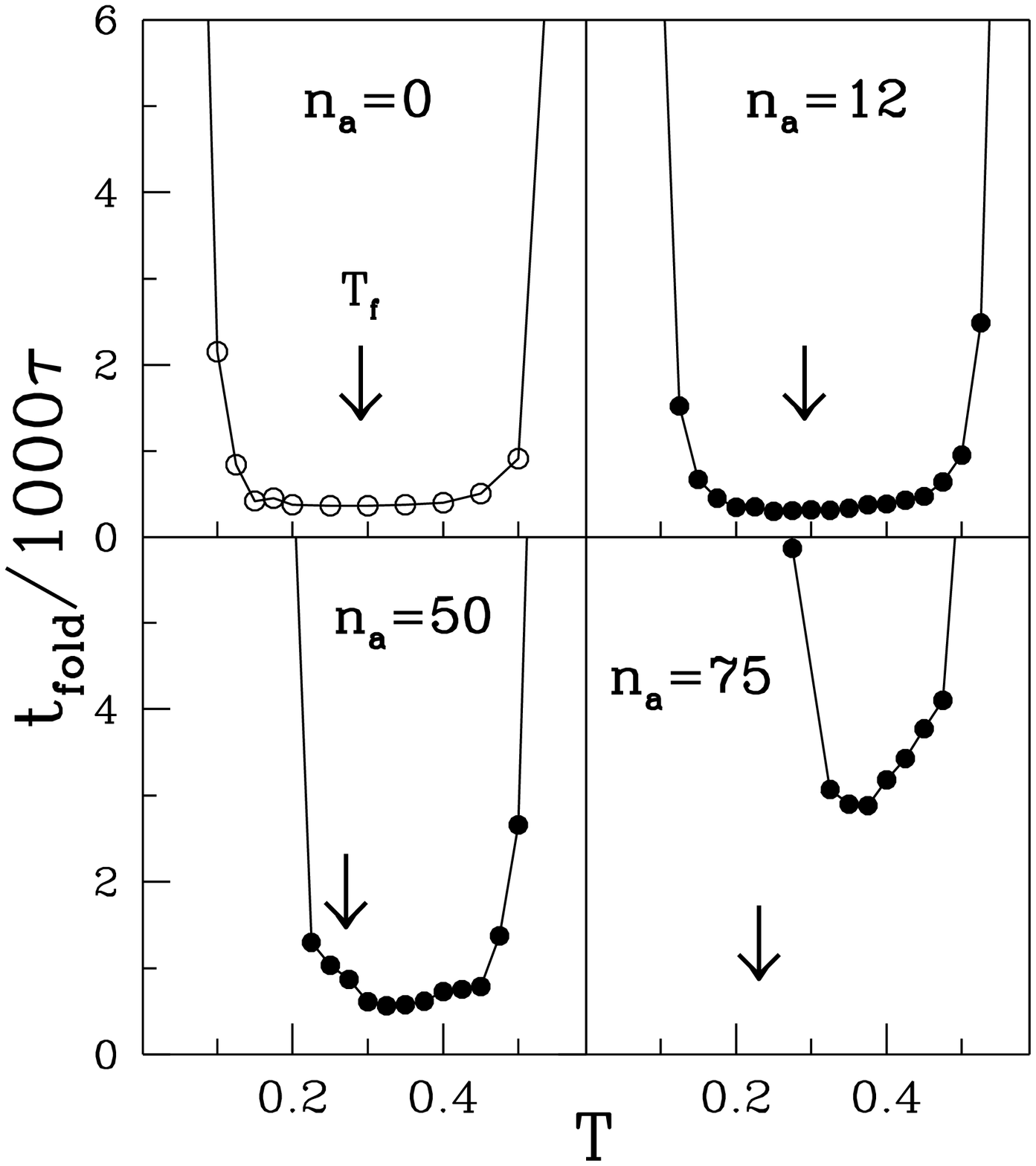}}
\vspace*{0.5cm}
\caption{ }
\end{figure}

\begin{figure}
\vspace*{-0.5cm}
\epsfxsize=7in
\centerline{\epsffile{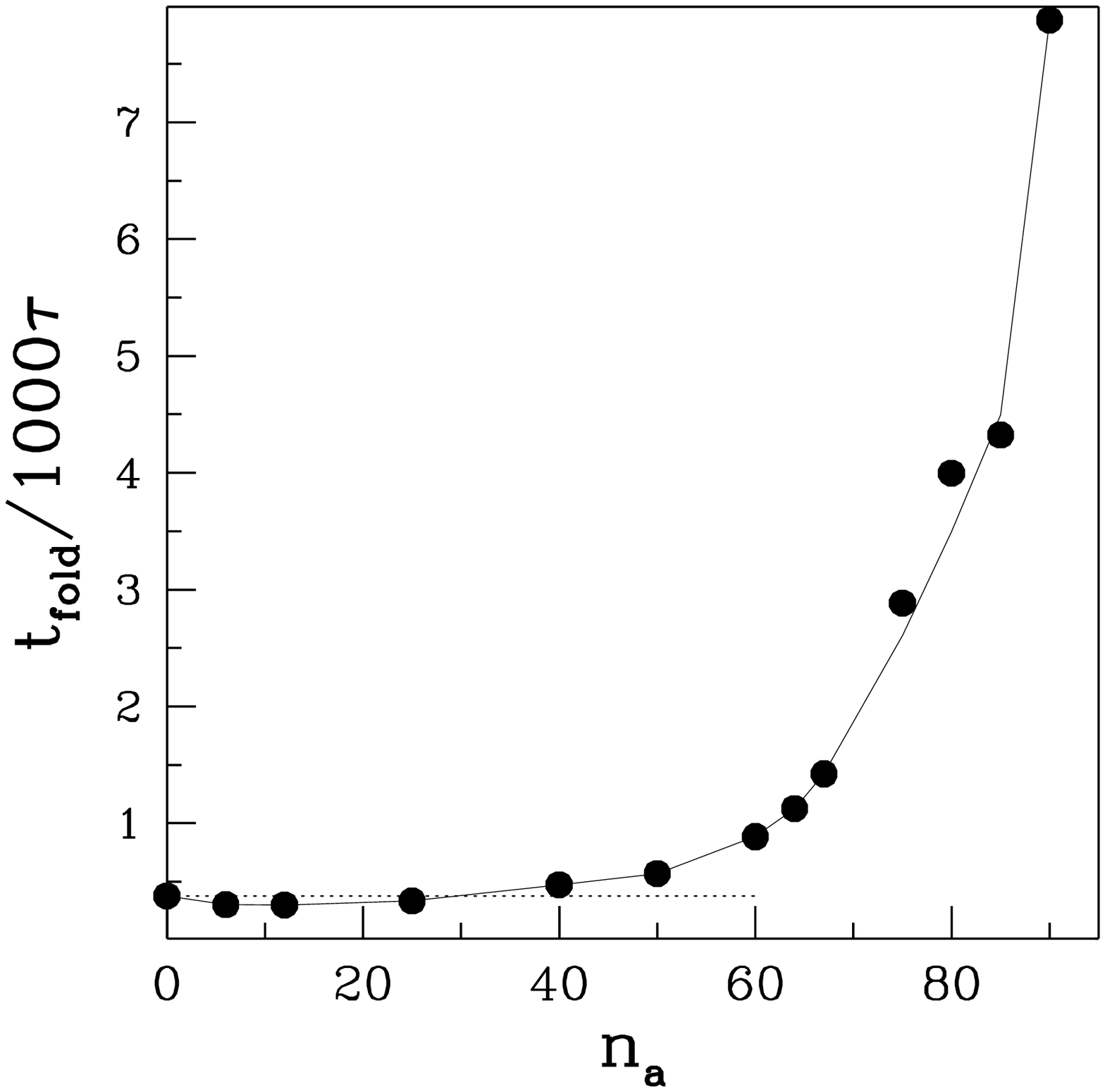}}
\vspace*{0.5cm}
\caption{ }
\end{figure}

\begin{figure}
\vspace*{-0.5cm}
\epsfxsize=7in
\centerline{\epsffile{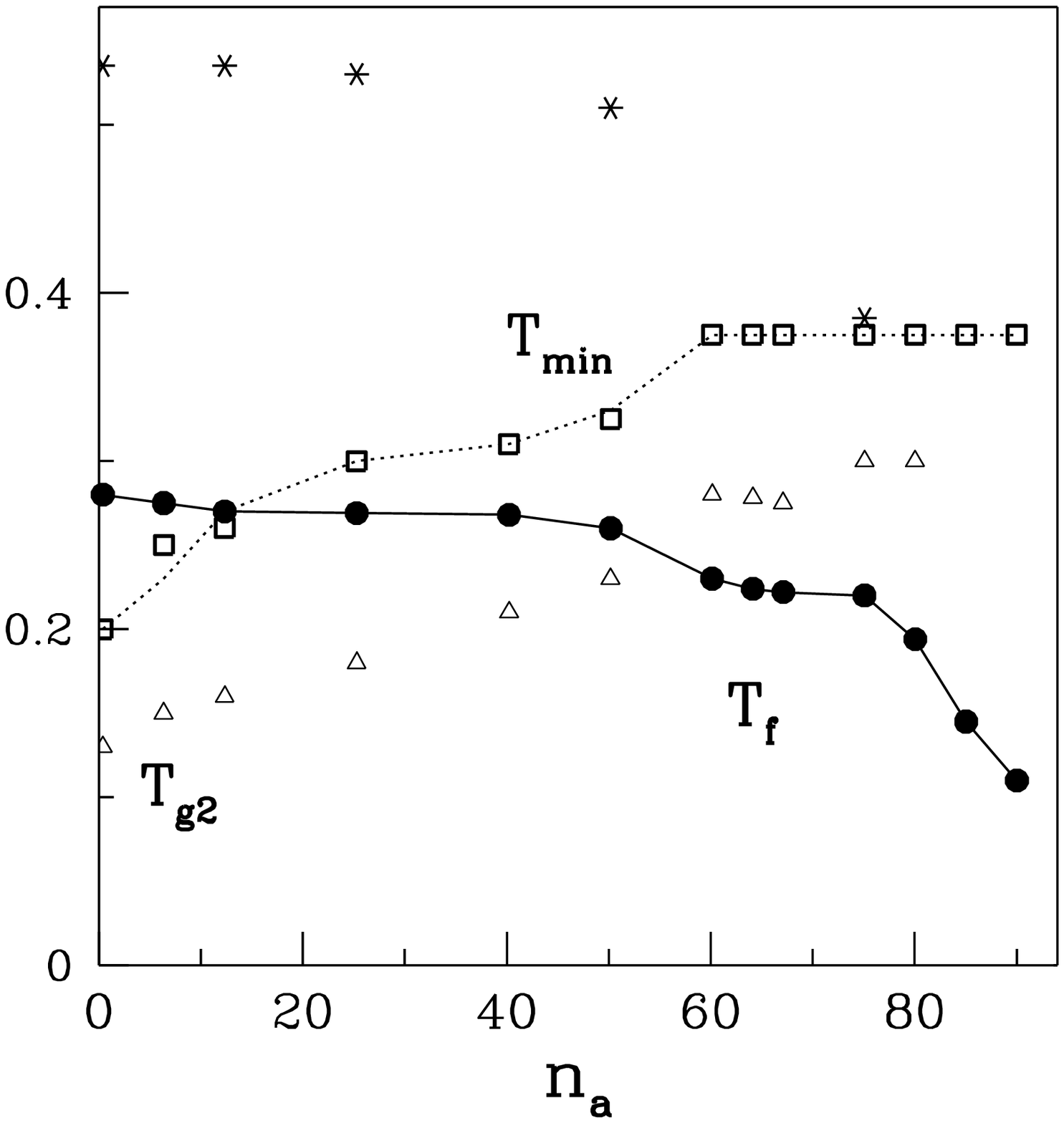}}
\vspace*{0.5cm}
\caption{ }
\end{figure}

\begin{figure}
\vspace*{-0.5cm}
\epsfxsize=7in
\centerline{\epsffile{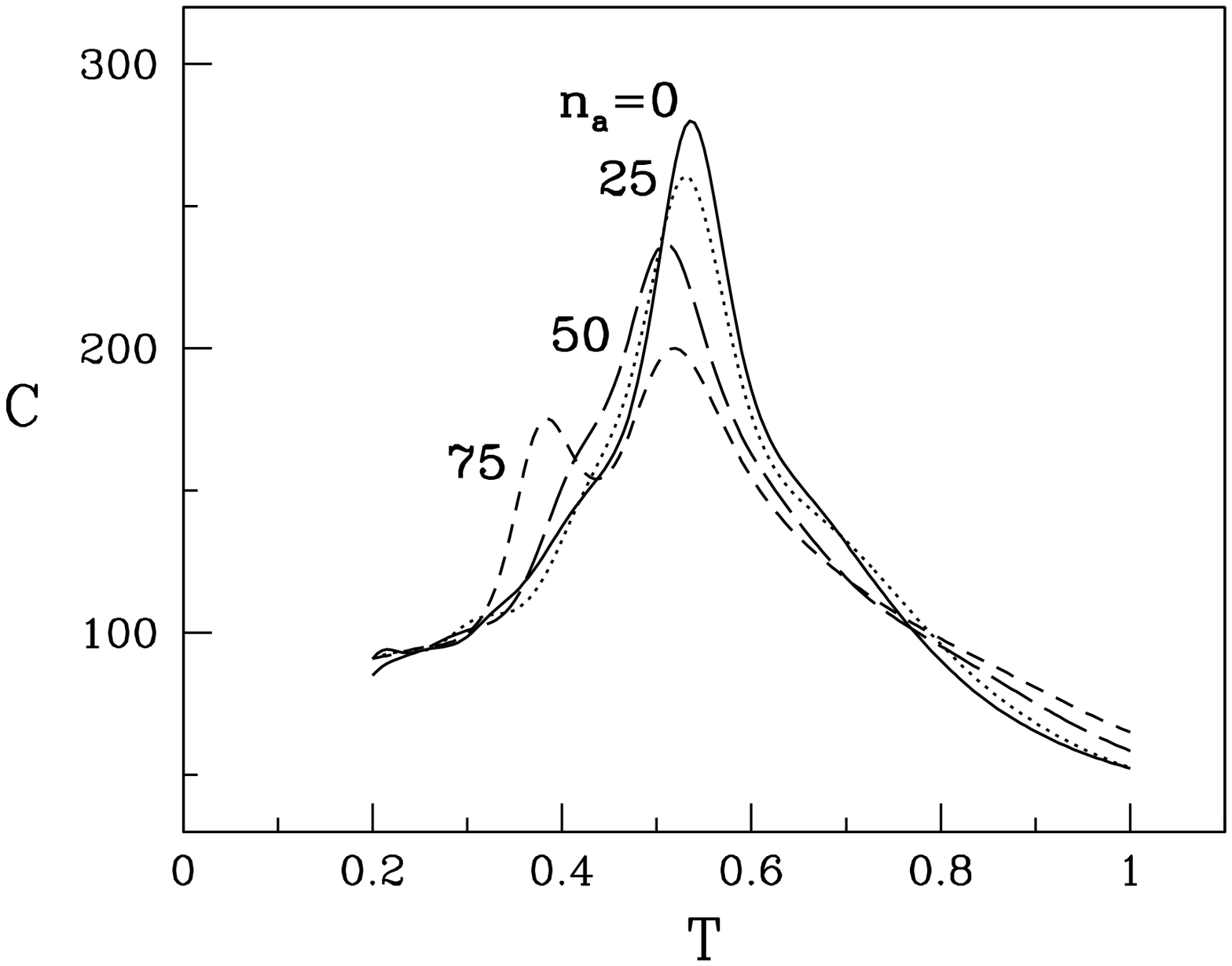}}
\vspace*{0.5cm}
\caption{ }
\end{figure}

\end{document}